\documentclass[fleqn,usenatbib]{mnras}

\usepackage{newtxtext,newtxmath}
\usepackage[T1]{fontenc}

\DeclareRobustCommand{\VAN}[3]{#2}
\let\VANthebibliography\thebibliography
\def\thebibliography{\DeclareRobustCommand{\VAN}[3]{##3}\VANthebibliography}

\usepackage{graphicx}	% Including figure files
\usepackage{amsmath}	% Advanced maths commands
\usepackage{amssymb}	% Extra maths symbols

\usepackage{caption}

\newcommand{\heii}{He~\textsc{ii}~}

\newcommand{\civ}{C~\textsc{iv}~}
\newcommand{\ciii}{C~\textsc{iii}]~}
\newcommand{\oiii}{O~\textsc{iii}]~}

\defcitealias{sax20}{S20}

\title[X-ray properties of \heii emitters]{X-ray properties of \heii $\mathbf{\lambda1640}$ emitting galaxies in VANDELS}

\author[A. Saxena et al.]{
A. Saxena,$^{1,2}$\thanks{E-mail: aayush.saxena@ucl.ac.uk}
L. Pentericci,$^{1}$
D. Schaerer,$^{3}$
R. Schneider,$^{1,4}$
R. Amorin,$^{5,6}$
\newauthor A. Bongiorno,$^{1}$
A. Calabr\`{o},$^{1}$
M. Castellano,$^{1}$
A. Cimatti,$^{7,8}$
F. Cullen,$^{9}$
A. Fontana,$^{1}$
\newauthor J. P. U. Fynbo,$^{10}$
N. Hathi,$^{11}$
D. J. McLeod,$^{9}$
M. Talia$^{12}$
and G. Zamorani$^{12}$
\\ \\
$^{1}$INAF -- Osservatorio Astronomico di Roma, via Frascati 33, 00078, Monteporzio Catone, Italy\\
$^{2}$Department of Physics and Astronomy, University College London, Gower Street, London WC1E 6BT, UK\\	
$^{3}$Observatoire de Gen\`{e}ve, Universit\'{e} de Gen\`{e}ve, 51 Ch. des Maillettes, 1290 Versoix, Switzerland\\
$^{4}$Dipartimento di Fisica, Sapienza Universit\`{a} di Roma, Piazzale Aldo Moro 5, 00185, Roma, Italy\\
$^{5}$Instituto de Investigaci\'{o}n Multidisciplinar en Ciencia y Tecnolog\'{i}a, Universidad de La Serena, Ra\'{u}l Bitr\'{a}n 1305, La Serena, Chile\\
$^{6}$Departamento de F\'{i}sica y Astronom\'{i}a, Universidad de La Serena, Av. Juan Cisternas 1200 Norte, La Serena, Chile\\
$^{7}$University of Bologna, Department of Physics and Astronomy (DIFA), Via Gobetti 93/2, I-40129, Bologna, Italy\\
$^{8}$INAF - Osservatorio Astrofisico di Arcetri, Largo E. Fermi 5, 50125, Firenze, Italy\\
$^{9}$SUPA (Scottish Universities Physics Alliance), Institute for Astronomy, University of Edinburgh, Royal Observatory, EH9 3HJ Edinburgh, UK\\
$^{10}$Cosmic DAWN Center, Niels Bohr Institute, University of Copenhagen, Juliane Maries Vej 30, 2100 Copenhagen \O, Denmark\\
$^{11}$Space Telescope Science Institute, 3700 San Martin Drive, Baltimore, MD 21218, USA\\
$^{12}$INAF -- OAS Bologna, Via P. Gobetti 93/3, 40129, Bologna, Italy
}

\date{Accepted 2020 June 18. Received 2020 June 18; in original form 2020 March 30}

\pubyear{2020}

\begin{document}
\label{firstpage}
\pagerange{\pageref{firstpage}--\pageref{lastpage}}
\maketitle

\begin{abstract}
We explore X-ray emission from a sample of 18 \heii $\lambda1640$ emitting star-forming galaxies at $z\sim2.3-3.6$ from the VANDELS survey in the Chandra Deep Field South, to set constraints on the role of X-ray sources in powering the \heii emission. We find that 4 \heii emitters have tentative detections with $\textrm{S/N} \sim 2$ and have X-ray luminosities, $L_X = 1.5-4.9 \times 10^{41}$ erg~s$^{-1}$. The stacked luminosity of all 18 \heii emitters is $2.6 \times 10^{41}$ erg~s$^{-1}$, and that of a subset of 13 narrow \heii emitters (FHWM(He~\textsc{ii}) < 1000 km~s$^{-1}$) is $3.1 \times 10^{41}$ erg~s$^{-1}$. We also measure stacked $L_X$ for non-\heii emitters through bootstrapping of matched samples, and find $L_X = 2.5 \times 10^{41}$ erg~s$^{-1}$, which is not significantly different from $L_X$ measured for \heii emitters. The $L_X$ per star-formation rate for \heii emitters ($\log (L_X/\textrm{SFR}) \sim 40.0$) and non-emitters ($\log (L_X/\textrm{SFR}) \sim 39.9$) are also comparable and in line with the redshift evolution and metallicity dependence predicted by models. Due to the non-significant difference between the X-ray emission from galaxies with and without He~\textsc{ii}, we conclude that X-ray  binaries or weak or obscured AGNs are unlikely to be the dominant producers of \heii ionising photons in VANDELS star-forming galaxies at $z\sim3$. Given the comparable physical properties of both \heii emitters and non-emitters reported previously, alternative \heii ionising mechanisms such as localised low-metallicity stellar populations, Pop-III stars, etc. may need to be explored.
\end{abstract}


%%%%%%%%%%%%%%%%%%%%%%%%%%%%%%%%%%%%%%%%%%%%%%%%%%

%%%%%%%%%%%%%%%%% BODY OF PAPER %%%%%%%%%%%%%%%%%%

\section{Introduction}
\label{sec:introduction}
Low-mass star-forming galaxies are largely considered to be the key drivers of reionisation, a process through which the Universe made a phase transition from neutral to completely ionised by $z\sim6$ \citep{rob10, rob15, wis14, bou15}. With decreasing metallicities at higher redshifts \citep{hen13, ste14, amo17, san18, cul19}, galaxies in the early Universe should be capable of producing a large number of ionising photons ($E>13.6$ eV) and complete the process of reionisation by $z\sim6$ \citep{sta16}. The metal-free stars (the so-called Pop III stars) in these very early galaxies should have very high masses and temperatures \citep[e.g.][]{bro04, bro11}, resulting in the production of hard UV ionising fields that are capable of exciting high-ionisation emission lines, such as \heii $\lambda1640$, whose ionisation potential is $>54.4$ eV or $\lambda < 228$~\AA\ \citep{tum01, sch03, sca03}.

The number of known galaxies that show the high ionisation \heii emission line has been steadily growing across redshifts. In the local Universe ($z\sim0$), the \heii $\lambda 4868$ line is often seen in the spectra of low-mass star-forming galaxies and almost all of them are metal-poor \citep{gar91, gus00, izo04, shi12, keh15, keh18, ber16, sen17}. Rest-frame UV observations of some of these $z\sim0$ metal-poor galaxies have revealed the presence of both \heii $\lambda1640$ as well as \civ $\lambda 1540$ emission lines \citep{ber19, sen20}, reinforcing the idea of high ionisation due to massive, metal-poor stars. The samples of \heii emitting galaxies at high redshifts have increased too, primarily using lensing \citep{pat16, ber18} and large-area spectroscopic surveys \citep{cas13, nan19, sax20}, leading to detections of \heii emitting galaxies out to $z\sim4$.

Most of the broad \heii emission seen across redshifts can be explained primarily through winds driven by Wolf-Rayet (WR) stars \citep{sch96}. The WR origin in some broad \heii emitters (FWHM $>1000$ km~s$^{-1}$) has indeed been confirmed through the detection of WR `bumps' in the spectra of galaxies around the \heii and \civ emission lines \citep{bri08, keh11, shi12}. The inclusion of binary-star evolution in stellar population synthesis \citep{eld17, sta18} results in stars spending longer periods of time in the WR phase, and fits the observed \heii line better compared to single-star models \citep[e.g.][]{ste16}. However, not all broad \heii emitters, particularly those with low metallicities, may be directly connected with the presence of WR stars \citep[e.g.][]{shi12, keh15}. 

The picture becomes even more complicated when trying to explain the origin of the narrow \heii emission line (FWHM $<1000$ km~s$^{-1}$) \citep[e.g.][]{sta19}. Some stellar synthesis models including binary stars can reproduce the UV emission line ratios of \heii, \oiii and \ciii of galaxies that show \heii $\lambda 1640$ line observed at high redshifts. However, these models still under-predict the observed equivalent widths (EW) of the \heii line \citep{nan19, sax20}. Other physical mechanisms, such as strong shocks \citep{dop96, thu05, izo12}, stellar rotation mixing leading to higher effective temperatures \citep{szc15}, `stellar stripping' that results in the rejuvenation of old stars that provide extra \heii ionising photons \citep{got18, got19}, presence of metal-free Pop III stars \citep{sch03, cas13, vis17}, low-level AGN activity \citep[e.g.][]{mig19} and contribution from X-ray binaries (XRBs) \citep{gar91, sta15, keh15, sch19, sen20} have been proposed as possible explanations to account for the missing \heii ionising photons seen in star-forming galaxies.

XRBs are binary star systems where the production of X-rays is powered by mass transfer from the `donor' star to a very compact companion, such as a neutron star or black hole, which is called the `accretor'. The donor star can have a range of masses -- when the mass of the donor star is lower than the accretor, the system is referred to as a low-mass XRB. In cases where the donor star is massive, typically a O- or B-type star, the system is referred to as a high-mass XRB. The dominant sources of X-rays from young, star-forming galaxies at high redshifts are generally high-mass XRB systems \citep[e.g.][]{leh16}. Observations of X-ray emission from star-forming galaxies (at fixed star-formation rates) have revealed a strong metallicity-dependence of their X-ray luminosities. This means that the contribution from XRBs increases with decreasing metallicities \citep{bas13b, dou15, bro16, leh16, for19}. This metallicity dependence of XRBs has also been explored from a theoretical point of view \citep{lin10, fra13a, fra13b, mad17}. Especially in the early Universe, when the overall ages and metallicities of galaxies were lower and star-formation rates (SFRs) were higher, X-ray luminosities are also found to correlate strongly with the galaxy SFRs \citep{bas13a, leh16, air17}. This suggests that high-mass XRBs formed in star-forming regions within galaxies are the driving forces behind the observed X-ray luminosities of these galaxies. Since low metallicities and high-mass star-formation are also required to power nebular \heii emission, enhanced contribution from XRBs may offer an explanation to the missing \heii ionising photons problem \citep[e.g.][]{sch19}.

Building upon the new sample of \heii emitters at $z\sim2.2-5$ that was presented in \citet[][hereafter \citetalias{sax20}]{sax20}, in this paper we explore their X-ray properties, and compare them with those of the general star-forming galaxy population at similar redshifts. The layout of this paper is as follows: in Section \ref{sec:sample} we briefly outline the original sample of \heii emitting galaxies and their physical properties. In Section \ref{sec:xray} we introduce the X-ray data used in this study and present our methodology for X-ray photometry. In Section \ref{sec:results} we discuss the results of our X-ray analysis, and compare the X-ray properties of \heii emitters with samples of non-\heii emitters. In Section \ref{sec:discussion} we present a discussion of our results, and comment on whether X-ray sources play a dominant role in galaxies with \heii emission. Finally, we summarise the findings of this paper in Section \ref{sec:conclusions}.

Throughout this paper, we assume a $\Lambda$CDM cosmology with $\Omega_\textrm{m} = 0.3$ and H$_0 = 67.7$ km s$^{-1}$ Mpc$^{-1}$ taken from \citet{planck}, and use the AB magnitude system \citep{oke83}.  

\section{Sample of \heii emitters from VANDELS}
\label{sec:sample}
\subsection{Selection}
The sample of \heii emitting galaxies considered in this study was first presented in \citetalias{sax20}, and we refer the readers to this paper for the full description of sample selection, derived physical properties and analysis of both individual and stacked UV spectra. In this section we briefly summarise the key findings of \citetalias{sax20}. The galaxies were selected from VANDELS \citep{pen18, mcl18}, which is a recently completed deep VIMOS survey of the CANDELS CDFS and UDS fields \citep{gro11, koe11} carried out using the \emph{Very Large Telescope (VLT)}. We shortlisted a total of 50 star-forming galaxies over a redshift range $z=2.2-4.8$ that showed \heii emission in their spectra. Of these, 33 were classified as \emph{Bright} \heii emitters where the signal-to-noise ratio (S/N) of the \heii emission line was greater than 2.5, and 17 were classified as \emph{Faint} emitters with S/N(He \textsc{ii}) $<$ 2.5. Out of the 50 total shortlisted \heii emitters, 26 (19 \emph{Bright} and 7 \emph{Faint}) lie in the Chandra Deep Field South (CDFS) and 24 (14 \emph{Bright} and 10 \emph{Faint}) lie in UKIDSS Ultra Deep Survey (UDS) field.

\subsection{Physical properties}
Physical parameters such as stellar masses, star-formation rates (SFRs), and rest-frame absolute UV magnitudes (M$_{\textrm{UV}}$) were obtained by fitting spectral energy distribution (SED) templates to photometric points from broad-band filters at the spectroscopic redshift of each galaxy. The SED fits were performed using $Z = 0.2$ $Z_\odot$ metallicity versions of the standard \citet{bc03} models with redshifts fixed to the VANDELS spectroscopic redshift. The star-formation rates were corrected for dust adopting the \citet{cal00} dust attenuation law. The rest-frame magnitudes were calculated using a 200~\AA\ wide top-hat filter centred at 1500~\AA. We refer the readers to \citet{mcl18} for full details of the SED fitting techniques, model assumptions, and derived physical parameters for VANDELS sources.

Overall, \citetalias{sax20} found that galaxies that show \heii emission have comparable stellar masses, star-formation rates and UV magnitudes, to similarly selected VANDELS galaxies with no \heii emission over the same redshift range. \citetalias{sax20} reported that the stellar mass range of \heii emitters is $ \log_{10} \textrm{M}_\star  = 8.8 - 10.7$ M$_\odot$, the UV-corrected star-formation rate (SFR) range is $ \log_{10}(\textrm{SFR}) = 0.7 - 2.3$ M$_\odot \textrm{~yr}^{-1}$ and the absolute UV magnitude range is M$_{\textrm{UV}} = -21.9$ to $-19.2$. Two sample Kolmogorov-Smirnov (KS) tests showed that the physical properties of \heii emitters are not significantly different from those that do not show \heii emission in their spectra.

Next, UV emission line ratio diagnostics (\heii $\lambda1640$, \oiii $\lambda1666$, \ciii $\lambda1909$) were used to study the underlying physical conditions in star-forming galaxies that show \heii emission. Line ratios from both single-star models \citep{gut16} and binary-star models \citep{xia18} were used for this analysis. The comparison with models was performed using emission line ratios determined from individual galaxy spectra where the relevant UV lines were detected at high enough S/N, as well as stacks of spectra. In total, three additional stacked spectra were produced: (a) a stack of all \emph{Faint} \heii emitters, (b) stack of galaxies in the \emph{Bright} sample that show narrow \heii (FWHM < 1000 km~s$^{-1}$), and (c) stack of galaxies in the \emph{Bright} sample that show broad \heii (FWHM > 1000 km~s$^{-1}$). 

From comparing the line ratios, \citetalias{sax20} found that individual \heii emitters (with detections of other UV lines) largely favour sub-solar stellar metallicities and low stellar ages. From line ratios of stacked spectra, \citetalias{sax20} inferred that the stacks of faint, and bright and narrow \heii emitters favour lower metallicities compared to the line ratios from the stack of bright and broad \heii emitters. This is in line with predictions based on \heii ionising photons being produced in Wolf-Rayet (WR) stars \citep[see][for example]{shi12} -- higher metallicity stellar populations have more stars in the WR phase, which give rise to broad \heii emission lines due to strong stellar winds. 

For individual galaxies with bright \heii emission, as well as for the stack of faint and narrow \heii emitters, \citetalias{sax20} found that although binary-star models do a reasonably good job at reproducing the line ratios, they under-predict the \heii EWs. This means that these models are unable to produce the number of \heii ionising photons required to power the observed emission line strengths, and additional sources of ionising photons may be required. \citetalias{sax20} argued that there are several mechanisms that could be producing these missing photons, including sub-dominant AGN, stripped stars and/or X-ray binaries (XRBs), particularly the high-mass XRBs, as previously mentioned.

An effective way to investigate the impact of sub-dominant AGN or enhanced contribution from XRBs is to study the X-ray emission from \heii emitting galaxies. In Section \ref{sec:xray} we describe the available X-ray data and our X-ray photometry methodology.

\subsection{This work - the CDFS sample}
In this study, we focus on the \heii emitting galaxies in the CDFS field. Since the primary goal of this study is to measure X-ray fluxes from \heii emitters, access to ultra-deep X-ray data is essential. Therefore, we have chosen to restrict this analysis to the CDFS field, owing to the availability of \emph{Chandra} data with a total of 7 Ms of exposure time. Although \emph{Chandra} data is also available in the UDS field\footnote{\url{http://www.mpe.mpg.de/XraySurveys/XUDS/}}, the effective exposure time of the data available in UDS is $\sim600$ ks. This is quite shallow compared to data in the CDFS field and to detect faint star-forming galaxies at high redshifts, the depths reached by 600 ks of exposure time will not be sufficient. More details on X-ray data are given in the following section.

There are a total of 26 \heii emitters from \citetalias{sax20} (both \emph{Bright} and \emph{Faint} sources) that lie in the CDFS field. Out of these, 21 lie within the X-ray image footprint with high effective exposure times. We then cross-match the positions of \heii emitters with the CDFS 7 Ms source catalogue from \citet{cdfs7ms}, using a radius of 2 arcseconds. \citet{mag20} showed that the CDFS catalogue is complete down to X-ray luminosities of $10^{42}$ erg~s$^{-1}$ at $z\sim3$, and above these luminosities only AGN are found. Therefore, all sources that have a counterpart in the CDFS source catalogue are likely to be X-ray AGN. We do not find any matches between the \heii emitters and sources in the CDFS catalogue. 

Of the 21 sources within the X-ray footprint, 3 were classified as potential AGN by \citetalias{sax20} owing to the presence of strong \civ emission in their spectra. Interestingly, these 3 possible AGN are also not detected in the CDFS 7Ms catalogue. However, to be consistent with \citetalias{sax20} we take a conservative approach and remove these three sources from our sample.

The final sample, therefore, consists of 18 \heii emitters from \citetalias{sax20}. Out of these, 12 are \emph{Bright} \heii emitters, and 6 are \emph{Faint} \heii emitters. Based on the width of the \heii line, 13 have narrow (FWHM $< 1000$ km~s$^{-1}$) and 5 have broad (FWHM $>1000$ km~s$^{-1}$) \heii lines. The rest-frame UV spectra of \heii emitting galaxies can be found in \citetalias{sax20}. A breakdown of the number of sources and their classification based on their \heii line properties is given in Table \ref{tab:sample}.
\begin{table}
    \centering
    \caption{Number of sources and breakdown in terms of \heii line properties for sources in CDFS.}
    \begin{tabular}{l c r}
    \hline
    Class  & Property & Number \\
    \hline \hline
    CDFS X-ray footprint & All \heii & 21 \\
                         & Excluding AGN & 18 \\
    \hline
    \emph{Bright} & S/N(He \textsc{ii}) > 2.5 & 12 \\
    \emph{Faint} & S/N(He \textsc{ii}) < 2.5 & 6 \\
    \hline
    \emph{Narrow} & FWHM(He \textsc{ii}) < 1000 km~s$^{-1}$ & 13 \\
    \emph{Broad} & FWHM(He \textsc{ii}) > 1000 km~s$^{-1}$ & 5 \\
    \hline
    \end{tabular}
    \label{tab:sample}
\end{table}

\section{X-ray analysis}
\label{sec:xray}
\subsection{Data}
We use X-ray data from the \emph{Chandra X-ray Observatory} in CDFS, which has a total of 7 Ms of exposure time covering an area of $\sim 485$ arcmin$^2$ collected over a period of more than a decade \citep{cdfs7ms}\footnote{The images and catalogues are publicly available at \url{http://personal.psu.edu/wnb3/cdfs/cdfs-chandra.html}}, making it the deepest X-ray data set in any extragalactic field. Additional data products in the CDFS include the effective exposure map and the PSF map, which are used for aperture photometry. More details about the data reduction and products in the CDFS field that have been used in this study can be found in \citet{gia19}.  

\subsection{X-ray photometry of \heii emitters}
\label{sec:xray_phot}
To estimate X-ray fluxes for the 18 \heii emitters within the CDFS footprint, we use the $0.8-3$ keV band image because of two reasons. First, as \citet{gia19} showed, using the $0.8-3$ keV image instead of the standard soft X-ray band of $0.5-2$ keV results in higher number of counts recovered due to the higher transmissivity of the $0.8-3$ keV band. Second, the redshift distribution of the sources in this study is such that the $0.8-3$ keV band comes closest to rest-frame energy range of $2-10$ keV, upon which the analysis of this paper as well as several other observational studies that will be used for comparison are based. Therefore, the uncertainties arising from the application of $k$-corrections are minimised. 

We measure the X-ray flux of \heii emitters by performing aperture photometry using \textsc{photutils} \citep{photutils} at the RA and Dec of each source, taken from the VANDELS catalogue. Our methodology to measure the source and background counts for individual sources is as follows. We place a circular aperture encompassing the effective Chandra PSF (median diameter of 3.0$''$) at the positions of each \heii emitter to measure the total number of counts from the source. To measure the local background, we place a circular annulus with inner radius of $10''$ and outer radius of $20''$, centred on the same position as the circular aperture. Within the annulus, we mask pixels that are brighter than $4\sigma$ -- a relatively conservative value -- so as not to overestimate the background. Within the circular aperture and the annulus, we measure the total number of counts from the source and the background, $C_\textrm{gal}$ and $C_\textrm{bkg}$, the area encompassed, $A_\textrm{gal}$ and $A_\textrm{bkg}$ (pixel$^2$), and the effective exposure times, $t_\textrm{gal}$ and $t_\textrm{bkg}$ (seconds), respectively. We follow \citet{for19} and calculate the background subtracted counts as 
\begin{equation}
	C_\textrm{bkgsub} = C_\textrm{gal} - C_\textrm{bkg} \times \left(\frac{A_\textrm{gal}\times t_\textrm{gal}}{A_\textrm{bkg}\times t_\textrm{bkg}}\right)
\end{equation}

Since the counts from individual galaxies at these redshifts are expected to be low, we use \citet{geh86} approximation to establish confidence limits for a Poissonian distribution, which is standard practice when calculating errors in cases of low photon counts.

To convert from background subtracted counts in the $0.8-3$ keV band to X-ray flux in the standard $2-10$ keV band, which was selected to facilitate comparison with other similar studies, we must assume a spectral model to calculate the effective photon energy ($E_\textrm{eff}$) in the band and the appropriate $k$-correction ($k_\textrm{corr}$). In line with similar studies in the literature \citep[e.g.][]{bro16}, we assume a model with an un-obscured power-law spectrum with photon index $\Gamma=2.0$ and a galactic extinction value of $5\times 10^{20}$ cm$^{-2}$ \citep{voo12}, which is the average value observed for star-forming galaxies at high redshifts inferred from cosmological simulations. We then use \textsc{pimms}\footnote{\url{https://heasarc.gsfc.nasa.gov/docs/software/tools/pimms.html}} to calculate $E_\textrm{eff}$ required to convert counts in the observed frame $0.8-3$ keV band to fluxes in the observed frame $2-10$ keV energy range. We calculate X-ray fluxes in the $2-10$ keV range ($F_{\textrm{2-10 keV}}$) by dividing background subtracted counts by the effective exposure time ($t_\textrm{gal}$) and multiplying with $E_\textrm{eff}$, giving
\begin{equation}
	F_{\textrm{2-10 keV}} = \frac{C_\textrm{bkgsub}}{t_\textrm{gal}} \times E_{\textrm{eff}}
\end{equation}

To finally calculate rest-frame X-ray luminosities, we use luminosity distances, $D_L$, determined from the systemic redshift of sources given in \citetalias{sax20} and apply the $k$-correction, $k_\textrm{corr} = (1+z)^{\Gamma-2.0}$. Therefore, the rest-frame X-ray luminosities in the $2-10$ keV band ($L_{\textrm{2-10 keV}}$) are calculated as
\begin{equation}
	L_{\textrm{2-10 keV}} = F_{\textrm{2-10 keV}} \times 4\pi D_L^2 k_\textrm{corr}
\end{equation}

In Section \ref{sec:xray-ind} we present the X-ray properties of the individual sources that were detected with relatively high S/N.

\subsection{Stacking}
To boost the S/N of X-ray emission, we perform stacking analysis, where the stacked X-ray luminosity of $N$ sources is calculated as
\begin{equation}
	L_{X}^{\textrm{stack}} = \frac{1}{N} \sum\limits_{i}^N F_{X,i} \times 4\pi D_{L,i}^2 k_\textrm{corr}
\end{equation}

As shown by \citet{for19}, the above mentioned approximation to stacking works for galaxies that have similar $L_X$. Since in this study we are probing galaxies with similar physical properties such as redshifts, SFRs and masses, and the luminosities that go into the stack are weighted by the effective exposure time, we do not expect large inaccuracies in the stacked luminosity measured in this way. The errors on luminosity of each source are determined from the errors on the counts, which are then added in quadrature during stacking to obtain errors on the final stacked luminosities measured.

The stacking is performed on two samples of \heii emitters. The first sample includes all 18 \heii emitters, and the second sample only includes the 13 sources classified as \emph{Narrow} \heii emitters (FWHM < 1000 km~s$^{-1}$. The additional sub-sample of only narrow \heii emitters is created because as mentioned earlier, explaining the origin of the narrow \heii emission line is of particular interest in the context of contribution from X-ray sources. In Section \ref{sec:xray-ind} we also present the X-ray properties of the stacked \heii samples.

\subsection{Comparison samples of non-\heii emitters}
To understand the impact of X-ray sources in \heii emitting galaxies, we must compare the X-ray properties of \heii emitters with those of non-\heii emitters in VANDELS with similar properties such as redshifts, SFRs and stellar masses. For the two samples of \heii emitters considered in this study, we create two sub-samples of non-\heii emitters that have comparable physical properties to each \heii sample.
\begin{figure*}
    \centering
    \includegraphics[scale=0.75]{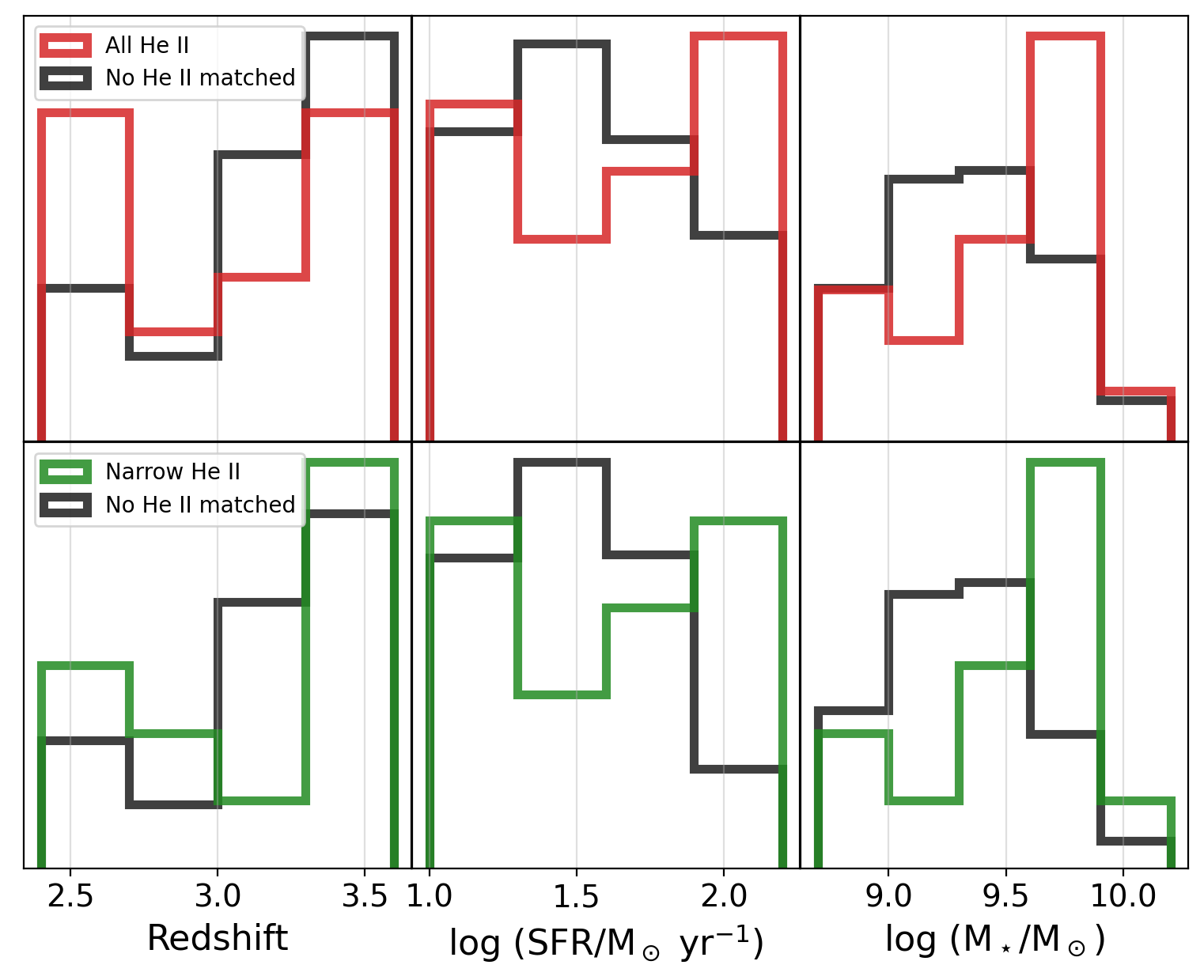}
    \caption{Normalised histograms of redshifts (left), star-formation rates (middle) and stellar masses (right) of the subsets of all \heii emitters (top panels) and narrow \heii emitters (bottom panels) compared with their respective matched parent samples. The matched samples of galaxies with no \heii have been created separately for the two sub-samples of \heii emitters to enable a more accurate comparison.}
    \label{fig:sfr-comparison}
\end{figure*}

To build a comparison sample for all \heii emitters, we select only those galaxies from VANDELS that lie in the redshift range $2.48 < z < 3.60$, have SFRs in the range $1.03 < \log (\textrm{SFR}/\textrm{M}_{\odot}\textrm{yr}^{-1}) < 2.19$ and stellar masses in the range $8.78 < \log(\textrm{M}_\star/\textrm{M}_\odot) < 9.94$. To create the comparison sample for narrow \heii emitters, the physical properties restrictions are $2.48 < z < 3.60$, $1.03 < \log (\textrm{SFR}/\textrm{M}_{\odot}\textrm{yr}^{-1}) < 2.04$ and $8.78 < \log(\textrm{M}_\star/\textrm{M}_\odot) < 9.94$. In Figure \ref{fig:sfr-comparison} we show the normalised histograms of redshifts, SFRs and stellar masses for both classes of \heii emitters considered in this study, along with their respective comparison samples.

We then identify any strong X-ray sources, most likely X-ray AGN, in the sample of non-\heii emitters by matching their coordinates with the CDFS 7 Ms source catalogue from \citet{cdfs7ms}, using a radius of 2 arcseconds. All sources that have a counterpart in the CDFS source catalogue are likely to be AGN and are removed from the comparison sample. Finally, we only consider those sources that lie in a high effective exposure time region in the CDFS 7 Ms image. This results in a total of 318 galaxies with similar physical properties compared to all \heii emitters, and 295 galaxies with similar properties compared to narrow \heii emitters, that lie within the footprint of the CDFS 7 Ms image.

We ensure comparable effective exposure times in the comparison samples by matching the number of galaxies that are randomly drawn for X-ray photometry from the non-\heii emitting galaxy sample. Therefore, to compare with all \heii emitters, we randomly draw 18 galaxies from the corresponding comparison sample, and to compare with narrow \heii emitters, we draw 13 galaxies from its comparison sample. For the randomly drawn galaxies, X-ray photometry and stacking is performed in a similar fashion to that of \heii emitters (described in Section \ref{sec:xray_phot}). This process is bootstrapped, resulting in 500 independent samples for which stacked X-ray luminosities are calculated for the comparison samples of both all and narrow \heii emitters. The final stacked X-ray luminosities and the associated errors from the comparison samples are measured from the median and standard deviation of the 500 independent bootstrap iterations. 

\subsection{X-ray luminosity per SFR ($\mathbf{L_X}$/SFR)} 
An important quantity that is often used to parametrise the effect of XRBs, primarily the high-mass XRBs, in star-forming galaxies is the X-ray luminosity per unit star-formation rate ($L_X$/SFR). \citet{leh16} showed that for star-forming galaxies at $z>2$ with specific SFRs $>10^{-8}$ (SFR/M$_\star$), high-mass XRBs are the dominant contributors to the X-ray emissivity. These high-mass XRBs drive the scaling relation between $L_X$ and SFR, as they begin to form only few tens of Myr after a starburst event and therefore, closely trace the star-formation rates \citep[see][for example]{ant16}.

Since the galaxies in question in this study are all star-forming galaxies at $z>2$, we also calculate and compare this quantity for both \heii emitters and non emitters to capture the contribution of these high-mass XRBs. The dust-corrected SFRs for VANDELS sources are derived from multi-band spectral energy distribution (SED) fitting, as described in \citet{mcl18}. The SFRs for \heii emitters along with more details are given in \citetalias{sax20}. 

The stacked $L_X$/SFR is calculated by dividing the $L_X$ by the SFR for each galaxy that goes into the stack, and the errors are propagated from the X-ray luminosities and added in quadrature. For the purposes of this study we ignore the errors on SFRs, as due to the relatively low X-ray counts expected from the sources, the bulk of the error on $L_X$/SFR should come from the error on $L_X$.

\section{Results}
\label{sec:results}
In this section we present results from X-ray photometry of \heii emitters and compare these with results for the bootstrap analysis carried out on the sample of non emitters from VANDELS.

\subsection{X-ray counts and luminosities}
\label{sec:xray-ind}

\subsubsection{Individual detections}
\begin{figure*}
	\centering
	\begin{minipage}{0.49\textwidth}
	    \includegraphics[scale=0.48]{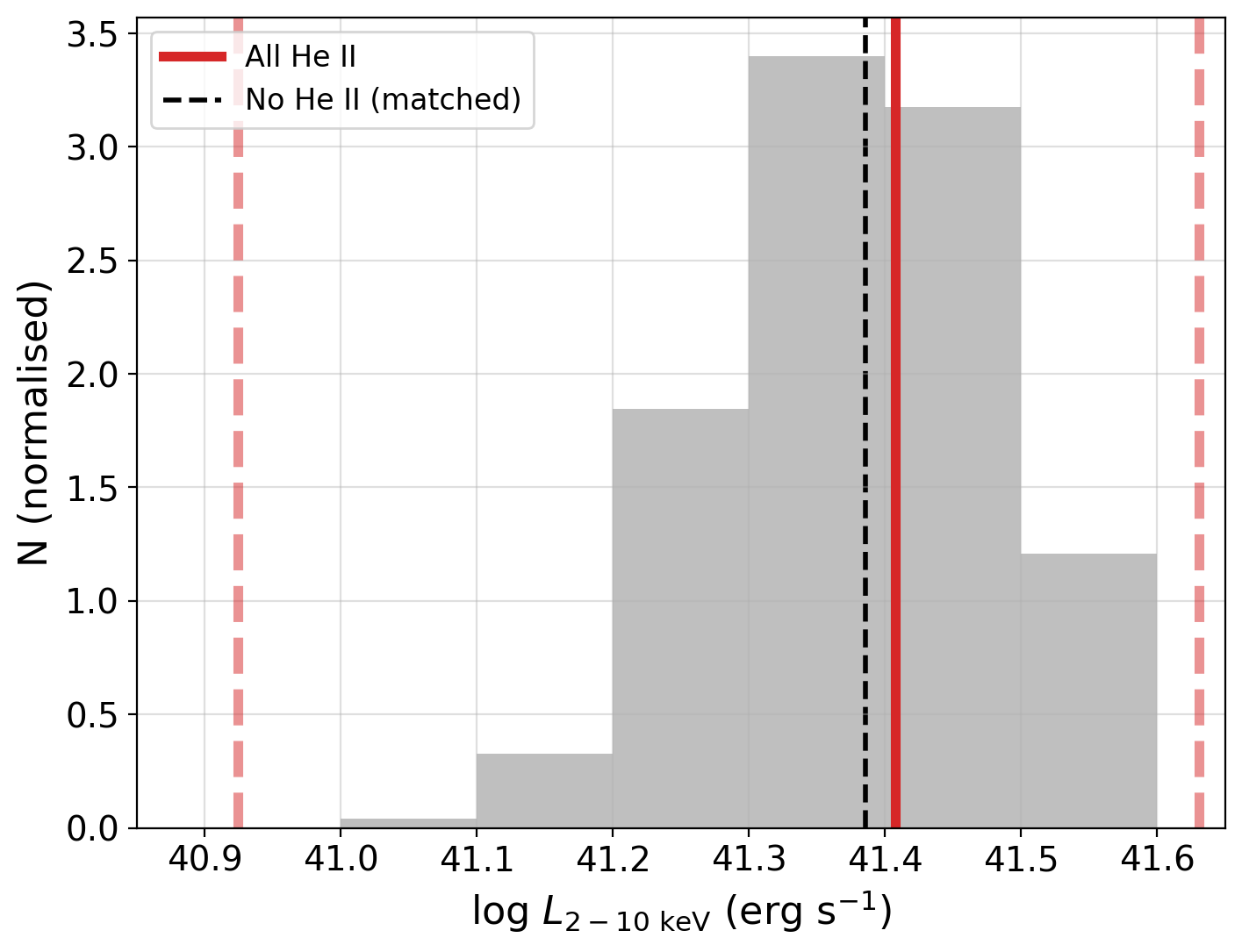}
	\end{minipage}
	\begin{minipage}{0.49\textwidth}
	    \includegraphics[scale=0.48]{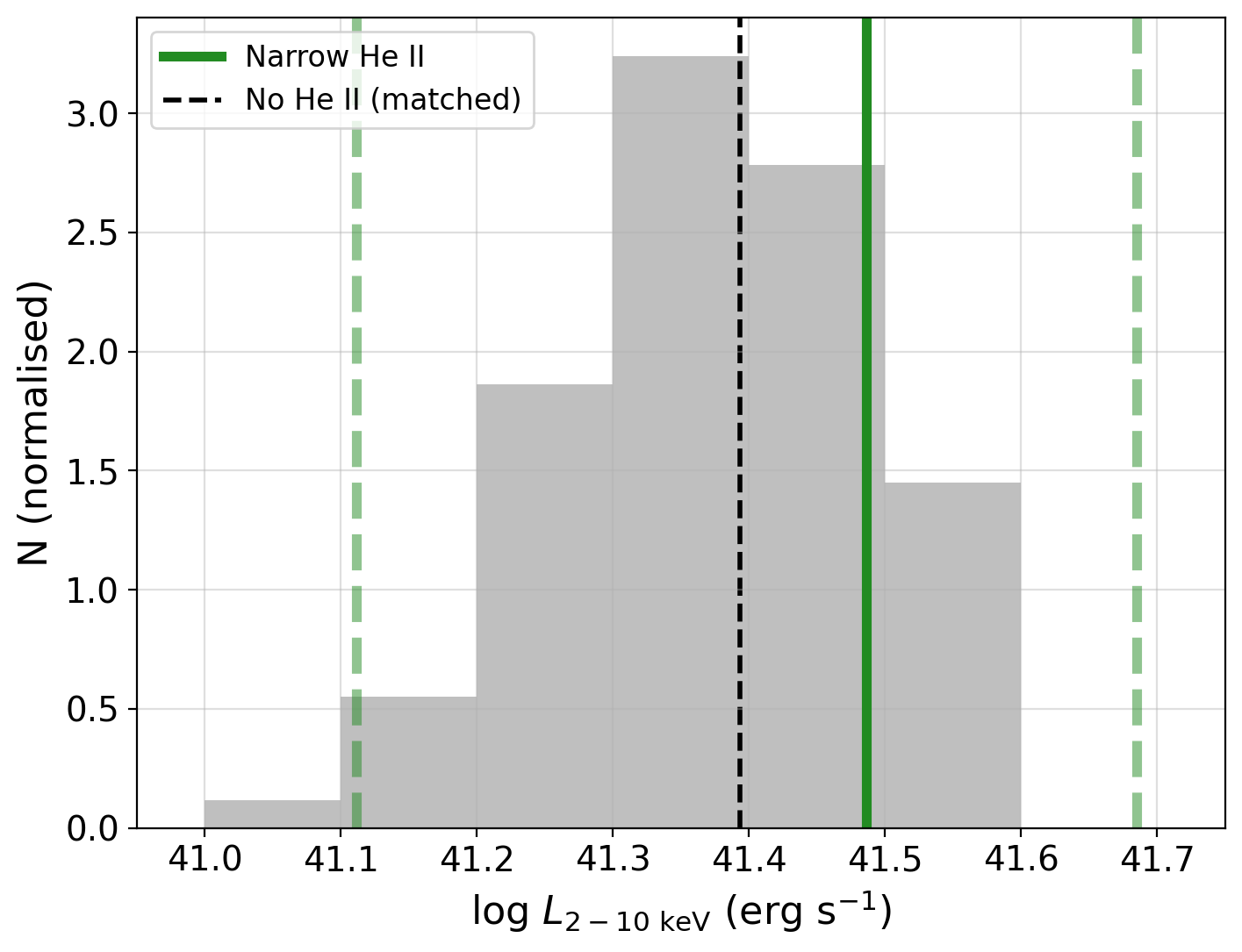}
	\end{minipage}
	\caption{The distribution of stacked X-ray luminosities in the $2-10$ keV band for the sample of all \heii emitters (left) and only narrow (FWHM < 1000 km~s$^{-1}$) \heii emitters (right). The distribution in grey shows stacked X-ray luminosity measured from a matched sample of galaxies with no \heii emission using bootstrapping in each panel. The dashed lines indicate the upper and lower errors on the measured luminosity of the sample of \heii emitters. Although narrow \heii emitters show higher X-ray luminosities in comparison, the Z-scores and P-values derived from comparing the stacked luminosities of both classes of \heii emitters with the distribution of non-\heii emitters shows that the difference between them is not statistically significant at $3\sigma$. This shows that there is no clear X-ray excess in \heii emitting galaxies.}
	\label{fig:counts}
\end{figure*}

We find that 4 out of 18 \heii emitters have counts with S/N $\ge 2$ to enable relatively reliable measurements of X-ray fluxes and luminosities. These sources were visually inspected to ensure that the emission is real, and not simply a distribution of noise peaks within the aperture. The background subtracted counts for these sources in the observed energy band $0.8-3$ keV band range from $7.9-14.4$, translating into luminosities in the rest-frame $2-10$ keV band of $L_X = 3.1-5.6\times 10^{41}$ erg~s$^{-1}$ (where $X=2-10$ keV). We note that of these 4 sources, 3 were classified as \emph{Bright} \heii emitters and one as \emph{Faint} \heii emitter by \citetalias{sax20} and interestingly, all four sources show narrow \heii emission lines (FWHM $< 1000$ km~s$^{-1}$). The X-ray properties of these individually detected sources along with the measured \heii luminosities from \citetalias{sax20} are given in Table \ref{tab:individual}.

Based on the X-ray luminosities of these individually detected sources, we can calculate the \heii ionising photon production rate per X-ray luminosity \citep[see][for example]{keh18, sch19}. To do that, we calculate the \heii ionising photon flux, Q(He \textsc{ii}), from the total \heii luminosity, L(He \textsc{ii}), by assuming Case B recombination and electron temperature $T_e = 30000$ K \citep[e.g.][]{sch19}. We find Q(He \textsc{ii}) values in the range $3.0-8.7 \times 10^{52}$ photons~s$^{-1}$. Dividing by the X-ray luminosities, we find $q = Q(\textrm{He \textsc{ii}})/L_X$ in the range $5.6-15.5\times10^{10}$ photons~erg$^{-1}$. These values are higher than what was found for the local dwarf galaxy I Zw 18 by \citet{sch19}, where $q \approx 1.0-3.4\times10^{10}$ photons~erg$^{-1}$. This suggests that the most X-ray bright sources in our sample of \heii emitters may have softer X-ray spectra than what has been observed for I Zw 18, assuming that all of the \heii emission is powered by the X-ray sources. The $q$ values measured for the individual sources are also given in Table \ref{tab:individual}.
\begin{table*}
	\centering
	\caption{X-ray properties of individually detected \heii emitting sources.}
	\begin{tabular}{l c c c c c c c r}
	\hline
	ID & Class & $z$ & $C_\textrm{bkgsub}$ & $L_X$ & $\log(\textrm{SFR})$ & $\log\frac{L_X}{\textrm{SFR}}$ & $L_{\textrm{He \textsc{ii}}}$ & $q = \frac{Q(\textrm{He \textsc{ii}})}{L_X}$ \\ 
	   &       &  & & \small{($10^{41}$ erg~s$^{-1}$)} & \small{($M_\odot$~yr$^{-1}$)} & & \small{($10^{41}$ erg~s$^{-1}$)} & \small{($10^{10}$ photons~erg$^{-1}$)} \\
	\hline \hline
	\vspace{3pt}
    CDFS23215 & B,N & 3.47 & 9.1 $\pm$ 3.0 & 5.6 $\pm$ 1.9 & 1.3 & $40.44^{+0.17}_{-0.27}$ & 4.9 $\pm$ 0.2 & 15.5 $\pm$ 4.5\\
  	\vspace{3pt}
  	CDFS113062 & B,N & 2.69 & 11.2 $\pm$ 3.3 & 3.1 $\pm$ 0.9 & 1.7 & $39.82^{+0.15}_{-0.22}$ & 2.5 $\pm$ 1.2 & 14.2 $\pm$ 3.0 \\
  	\vspace{3pt}
  	CDFS122687 & B,N & 2.64 & 14.4 $\pm$ 3.8 & 3.7 $\pm$ 1.0 & 1.9 & $39.60^{+0.13}_{-0.18}$ & 1.7 $\pm$ 0.8 & 8.2 $\pm$ 1.4 \\
  	CDFS10094 & F,N & 3.56 & 7.9 $\pm$ 2.9 & 4.6 $\pm$ 1.6 & 1.7 & $39.99^{+0.17}_{-0.29}$ & 1.5 $\pm$ 1.2 & 5.6 $\pm$ 2.7 \\
  	\hline
  	\end{tabular}\\
  	Class guide: B = \emph{Bright} (S/N(He \textsc{ii}) > 2.5), F = \emph{Faint} (S/N(He \textsc{ii}) < 2.5), N = \emph{Narrow} (FWHM(He \textsc{ii}) < 1000 km~s$^{-1}$), taken from \citetalias{sax20}.\\
  	\label{tab:individual}
 \end{table*}

\subsubsection{Stacks}
For the stack of all 18 \heii emitters (which include the 4 individually detected \heii emitters discussed above), the average background subtracted counts measured are $6.3 \pm 2.4$, and for the 13 narrow \heii emitters, the measured counts are $7.2 \pm 2.6$. The average background subtracted counts measured for non-\heii emitters using bootstrapping are $4.7 \pm 0.9$ per source, which are lower, but not significantly different than those measured for \heii emitters.
 \begin{table*}
    \centering
    \caption{X-ray properties of stacks.}
    \begin{tabular}{l c c c c c c r}
    \hline
	Stack & N & Total exp. & $\langle z \rangle$ & $\langle C_\textrm{bkgsub}\rangle$ & $\langle L_X \rangle$ & $\log \langle \textrm{SFR} \rangle$ & $\log\frac{\langle L_X \rangle}{\langle \textrm{SFR} \rangle}$ \\ 
	&   & \small{(Ms)}   &     & & \small{($10^{41}$ erg~s$^{-1}$)} & \small{($M_\odot$~yr$^{-1}$)} & \\ 
	\hline \hline
	All \heii & 18 & 126 & 3.04 & 6.2 $\pm$ 3.6 & 2.6 $\pm$ 1.7 & 1.7 & $40.01^{+0.26}_{-0.75}$\\
	\vspace{3pt}
	No \heii (matched)$^*$ & 18 & 126 & 3.15 & 4.3 $\pm$ 1.0 & 2.4 $\pm$ 0.7 & 1.6 & $39.95^{+0.12}_{-0.17}$\\
	Narrow \heii & 13 & 91 & 3.33 & 7.1 $\pm$ 3.8 & 3.1 $\pm$ 1.8 & 1.7 & $40.01^{+0.27}_{-0.82}$ \\
	No \heii (matched)$^*$ & 13 & 91 & 3.19 & 4.3 $\pm$ 1.1 & 2.5 $\pm$ 0.9 & 1.5 & $39.97^{+0.14}_{-0.22}$\\
	\hline
	\end{tabular}
	\label{tab:stacks}\\
	$^*$: errors measured on the stack of non-\heii emitters are from 500 bootstrap repetitions.

\end{table*}

We then compare the measured X-ray luminosities in the $2-10$ keV band from the stacks. We point out that the average redshifts of the two stacks are slightly different, which affects the calculation of the luminosity from the counts. We find that the stack of all \heii emitters with an average redshift of $\langle z \rangle =3.04$ has an X-ray luminosity $\langle L_X \rangle = 2.6 \pm 1.7 \times 10^{41}$ erg~s$^{-1}$. For the matched comparison sample of non-\heii emitters, the average redshift across 500 bootstrapped stacks is $\langle z \rangle =3.15$ and the average luminosity is $\langle L_X \rangle = 2.5 \pm 0.7 \times 10^{41}$ erg~s$^{-1}$. 

The stack of narrow \heii emitters with a higher average redshift of $\langle z \rangle =3.33$ have a slightly higher X-ray luminosity $\langle L_X \rangle = 3.1 \pm 1.8 \times 10^{41}$ erg~s$^{-1}$. The matched comparison sample of non-\heii emitters, with an average redshift of $\langle z \rangle =3.19$ has an X-ray luminosity $\langle L_X \rangle = 2.5 \pm 0.9 \times 10^{41}$ erg~s$^{-1}$. The X-ray properties of the stacks are given in Table \ref{tab:stacks}.

In Figure \ref{fig:counts} we mark the stacked X-ray luminosity of all \heii emitters (left panel) and narrow \heii emitters (right panel) against the distribution of X-ray luminosities measured from bootstrapping for their respective comparison samples of non-\heii emitters (grey histogram). The black dashed lines indicate the median luminosity inferred from bootstrapping of non-\heii emitters, and the dashed coloured lines mark the upper and lower X-ray luminosity confidence intervals for \heii emitters.

Our results show that the X-ray luminosities of \heii emitting galaxies are marginally higher than that of galaxies with no \heii emission. In particular, we find that the stack of narrow \heii emitters has the highest X-ray luminosity. However, we note that the X-ray measurements from \heii emitters are within $1\sigma$ of the distribution of X-ray luminosities from their respective comparison samples of non-\heii emitters. To calculate the statistical significance of the X-ray luminosities of \heii emitters, we calculate their Z-scores and P-values, which roughly gives the probability of a measurement being a statistical fluctuation from a given distribution. For the stacked luminosity of all \heii emitters, we find a Z-score of 0.40, giving a P-value of 0.355, indicating that there is a 35.5\% chance of this measurement being a statistical fluctuation and is not significantly different from the distribution of luminosities of non-\heii emitters. For the stack of narrow \heii emitters, we find a Z-score of 1.45 and a P-value of 0.073, indicating a statistical fluctuation probability of 7.4\%. Although the X-ray luminosity of narrow \heii emitters lies further away from the median, the inferred P-value still indicates that the difference is not statistically significant ($< 3\sigma$). 

From our X-ray measurements, we conclude that although \heii emitters, and narrow \heii emitters in particular, show marginally higher X-ray luminosities when compared to non-\heii emitting galaxies, the difference between the stacked X-ray luminosities of \heii emitting and non-emitting galaxy populations is not statistically significant. Therefore, within the statistical uncertainties presented with having small sample sizes, we do not find evidence of enhanced contribution from X-ray sources, presumably X-ray binaries or weak AGN, in galaxies that show the \heii emission line at $z\sim3$. We discuss this implication further in Section \ref{sec:discussion}.

\subsection{$\mathbf{L_X}$/SFR}
Before calculating $L_X$/SFR, we note that the median SFRs of the \heii emitting galaxies and non-emitting galaxies considered in this study are slightly different. For both stacks of \heii emitters, the median SFRs are $\log (\langle \textrm{SFR}) \rangle \sim 1.7$ M$_\odot$~yr$^{-1}$, which is higher than the median SFR of galaxies that do not show \heii with $\log(\langle \textrm{SFR} \rangle) \sim 1.5$ M$_\odot$~yr$^{-1}$. We note once again that the SFRs for all galaxies considered in this study are derived using multi-band SED fitting. For the stack of all \heii emitting galaxies, we calculate $\log(\langle L_X \rangle / \langle \textrm{SFR} \rangle)=40.03$ erg~s$^{-1}$/($M_\odot$~yr$^{-1}$), for narrow \heii emitters we calculate $\log(\langle L_X \rangle / \langle \textrm{SFR} \rangle)=40.11$ erg~s$^{-1}$/($M_\odot$~yr$^{-1}$), and for galaxies with no \heii emission, we calculate $\log(\langle L_X \rangle / \langle \textrm{SFR} \rangle)=39.93$ erg~s$^{-1}$/($M_\odot$~yr$^{-1}$). Similar to X-ray luminosities, we once again find that the $L_X$/SFR values of \heii emitters are marginally higher than non-emitters, but these measurements are not significantly different from each other. Therefore, we conclude that there is no clear excess of $L_X$/SFR in galaxies that show (narrow) \heii emission and those that do not.

\subsubsection{Redshift evolution}
\begin{figure}
	\centering
	\includegraphics[scale=0.44]{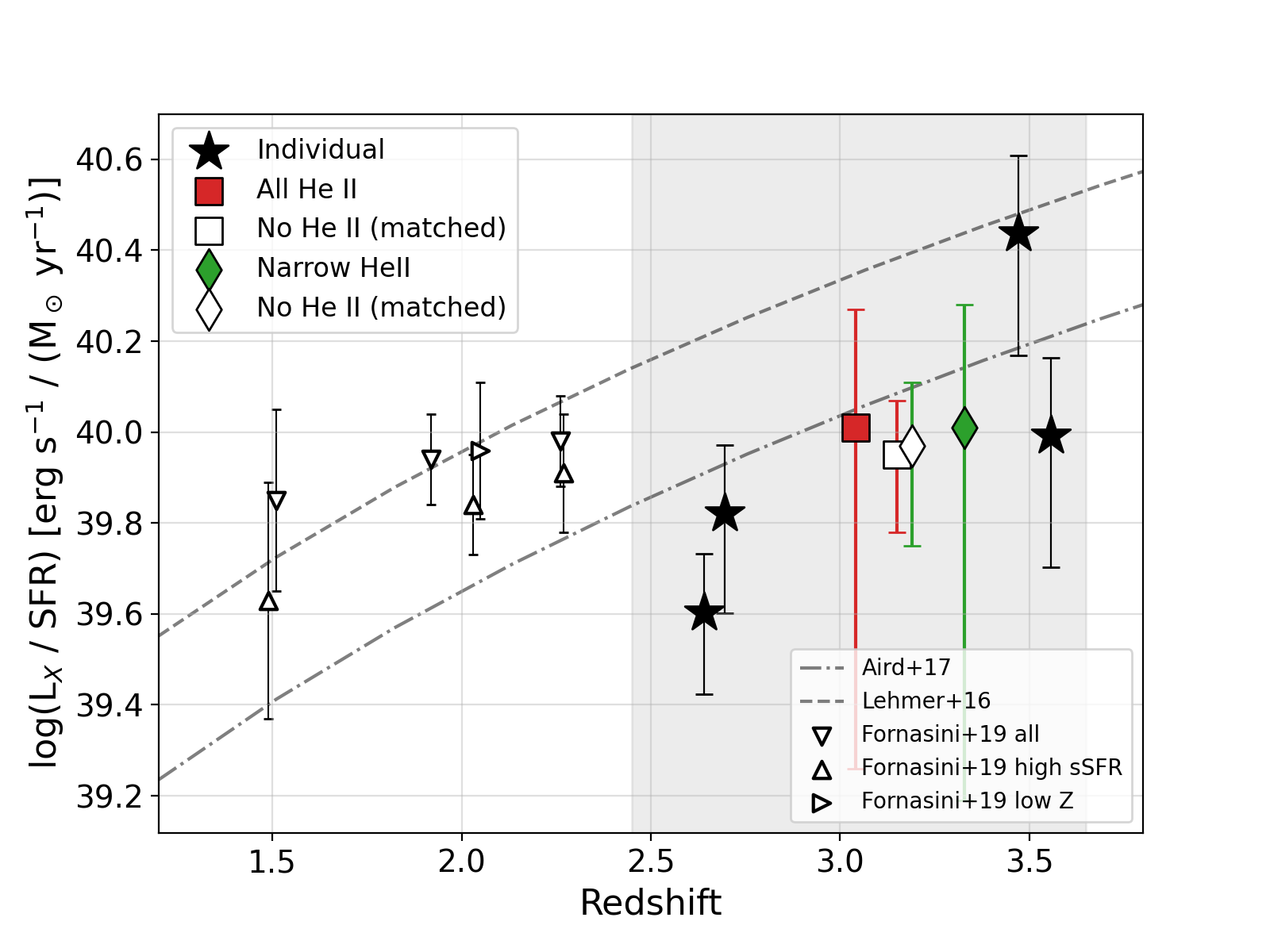}
	\caption{$L_X$/SFR versus redshift for star-forming galaxies at high redshifts. The coloured symbols show $L_X$/SFR measured in stacks, and black stars represent individual measurements in \heii emitters with the highest S/N in the X-ray image. The open symbols show measurements of lower redshift star-forming galaxies from \citet{for19}. The shaded region represents the range of redshift of sources that have been stacked in our measurements. We also show redshift evolution predicted by models from \citet{leh16} and \citet{air17}. We do not find a significant difference in $L_X$/SFR for galaxies with \heii and those without. Further, our measurements for both classes of galaxies are consistent with values found at $z\sim2$.}
	\label{fig:lx-sfr}
\end{figure}
To place our measurements of $L_X$/SFR for both \heii emitters and non-emitters within the general population of star-forming galaxies at high redshifts, we compare our measurements to those in the literature. We begin by looking at the redshift evolution of $L_X$/SFR inferred from samples of star-forming galaxies at $z\sim2$ from \citet{for19}. In Figure \ref{fig:lx-sfr} we show models predicting the redshift evolution of $L_X$/SFR from \citet{leh16} and \citet{air17}, the values measured by \citet{for19} at $z\sim2$, along with our measurements both for individually detected sources and stacks. The shaded region marks the redshift range probed in this study. We note that the redshift evolution models shown essentially capture the `X-ray main sequence' of star-formation, and have been calibrated using measurements at lower redshifts. The model predictions shown are normalised for star-formation rates of 20 $M_\odot$~yr$^{-1}$ to best match the observations from \citet{for19}.

We find that our measurement of $L_X$/SFR for individually detected sources and stacks of both \heii emitters and non emitters are consistent with what has been measured for star-forming galaxies at $z\sim1.5-2.5$, and in line with model predictions. Overall, we find little to no evolution in $L_X$/SFR between redshifts of 2 to 3. However, a proper study that captures the X-ray flux from the entire star-forming population in a systematic fashion is required to more accurately determine whether or not there is any redshift evolution out to $z\sim3$.

We note here that a key difference between the $L_X$/SFR determined for our sample and that of \citet{for19} is how the SFRs are measured. For our sources, we rely on SED derived SFRs using photometry at rest-frame UV to optical wavelengths, whereas the SFRs for a majority of sources in the \citet{for19} sample are derived using direct measurements of the H$\alpha$ emission line. Therefore, the timescales of the star-formation rates derived from the SED and the H$\alpha$ line would be different.

\subsubsection{Dependence on metallicity}
Several studies have explored the dependence of $L_X$/SFR on stellar metallicity for star-forming galaxies, both from theoretical \citep{fra13b, fra13a, mad17} and observational points of view \citep{bas13a, bas13b, bro16, for19}. Almost all evidence points towards a negative correlation between $L_X$/SFR and metallicity, both in the local and high-redshift Universe. This anti-correlation is driven primarily by the presence of higher mass black hole binaries at lower metallicities, that increases the contribution of high-mass XRBs to the overall $L_X$/SFR measured.
\begin{figure}
	\centering
	\includegraphics[scale=0.44]{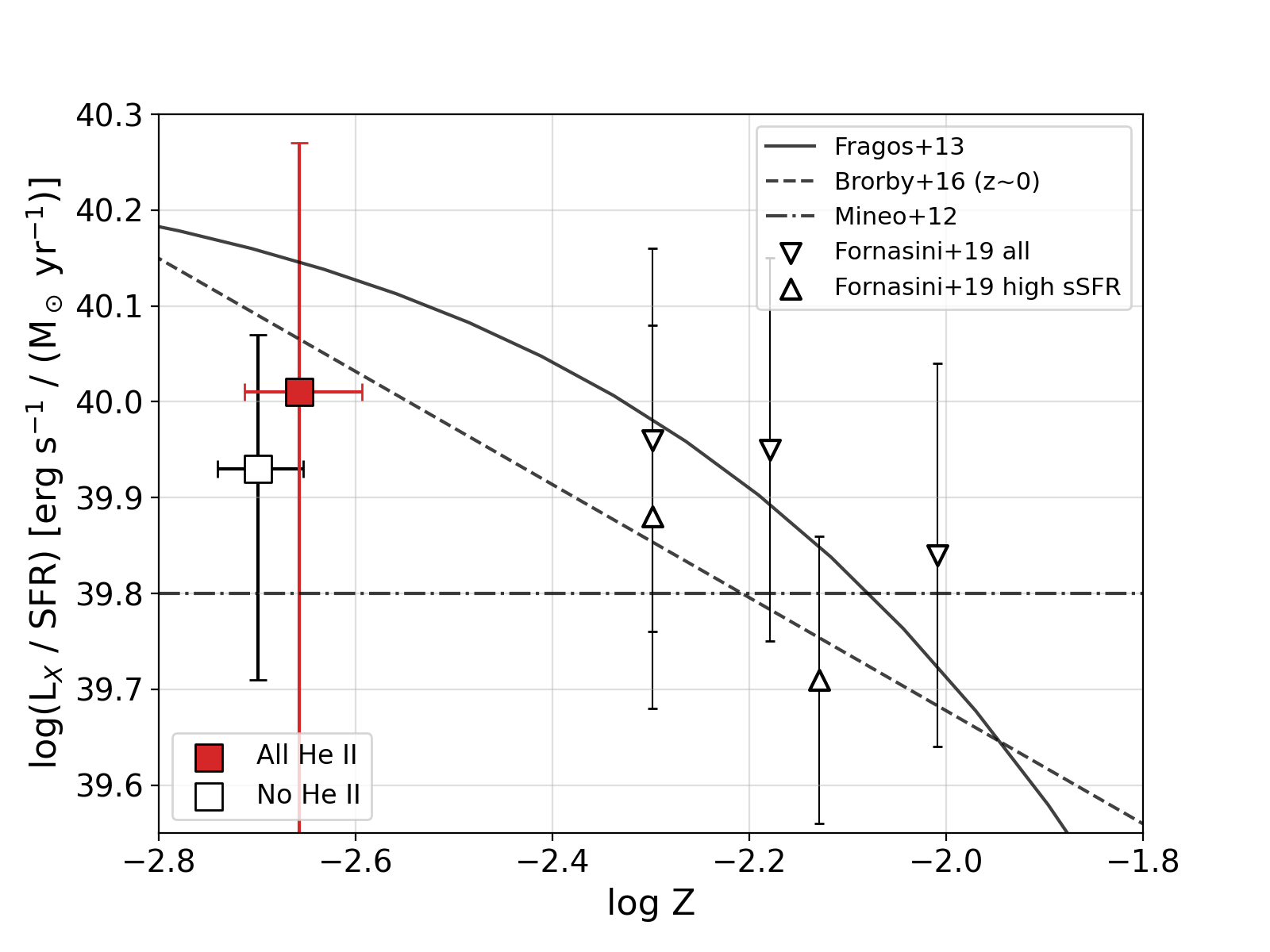}
	\caption{The dependence of $L_X$/SFR on stellar metallicity. The orange square shows the measurement for all \heii emitters, and red circle represents the non-\heii emitters. Also shown for comparison are observations from \citet{for19} for their samples of star-forming galaxies at $z\sim2$. Overlaid on top of the observations are predictions of XRB models from \citet{fra13b}, in addition to the local relation between $L_X$/SFR and metallicity observed by \citet{bro16}. The metallicities measured for both \heii emitters and non-emitters are taken from \citetalias{sax20}. Our measurements for both classes of galaxies at $z\sim3$ are comparable with observations at lower redshifts, and also in line with both model predictions and the $z\sim0$ relation. The assumed solar metallicity here is $\log~Z=-1.7$.}
	\label{fig:lx-metallicity}
\end{figure}

For our \heii and non-\heii emitting galaxies, we presented stellar metallicity measurements in \citetalias{sax20}, which were performed by fitting spectral features in the UV spectrum following the method of \citet{cul19}. To achieve high enough S/N to enable metallicity measurement, we only used stacks of all \heii emitters. Although such methods may not be as accurate as direct metallicity measurements from rest-frame optical emission lines, the stellar metallicities inferred can still provide valuable insights. In the context of predictions from models and previous observational evidence, we now compare whether our $L_X$/SFR measurements are in line with its dependence on metallicity that has been previously seen.

We once again compare our measurements with those of \citet{for19}. Note here that \citet{for19} use the gas-phase (O/H) ratios derived from spectroscopy as a proxy for stellar metallicity for their sample of star-forming galaxies. Since their metallicity measurements were made using rest-frame optical spectroscopy, they benefit from direct measurements of (O/H) ratios. For ease of comparison, we convert these (O/H) ratios to metal mass-fraction $Z$, using $Z=\textrm{(O/H)}*(\textrm{H}_\textrm{frac}/\textrm{O}_\textrm{frac})$, where $\textrm{H}_\textrm{frac}$ is the mass fraction of Hydrogen and $\textrm{O}_\textrm{frac}$ is the mass fraction of Oxygen. We find that assuming 40\% of O and 75\% of H are trapped in metals gives us consistent values when recovering the solar values for both (O/H) and $Z$. Our measurements along with observations from \citet{for19} are shown in Figure \ref{fig:lx-metallicity}. Also shown are the predictions from \citet{fra13b}, along with the best-fit power-law to data at $z\sim0$ from \citet{bro16} and the case of no metallicity dependence of $L_X$/SFR, as was reported by \citet{min12}.

Our measurements are in agreement with the metallicity dependence predicted by models and what has been reported in the literature. The $L_X$/SFR for \heii emitters is in line with the metallicity dependence predicted from models when compared to the \citet{for19} measurements made for galaxies with high specific SFRs. We also note that our measurements at $z\sim3$ are also consistent with the metallicity dependence of $L_X$/SFR measured in the local Universe by \citet{bro16}. Given the relatively large error bars on the stacked luminosities determined for galaxies in this work, our results are also consistent with a scenario where there is little to no evolution in the $L_X$/SFR with metallicity as reported by \citet{min12}. In a future study we aim to explore this metallicity dependence in more detail, extending the analysis to the full VANDELS sample of star-forming galaxies.

As \citet{sax20} noted, there is a slight caveat of the stellar metallicity measurement method from \citet{cul19} used in their work. With this method, the template fitting used to determine stellar metallicities from features in the rest-frame UV spectra assumes a constant star-formation history. This assumption is valid for averaging across the general star-forming galaxy population at high redshifts, but if galaxies (for example those with \heii) are very young, then their true metallicities may be higher than what is inferred using this method.

\section{Discussion}
\label{sec:discussion}
\subsection{No evidence of enhanced XRB contribution in \heii emitters}
\label{sec:xrb}
We find that the differences between $L_X$ and $L_X$/SFR of \heii emitters and non-emitters are not statistically significant. These results suggest that there is no excess X-ray emission, of whatever origin, in galaxies that show strong \heii emission in their spectra. As shown in \citetalias{sax20}, the metallicities measured for both \heii emitters and non-emitters are comparable too, in addition to physical properties such as stellar mass and star-formation rates. Since the $L_X$/SFR we measure for both classes of galaxies are also consistent with models and predictions for the general star-forming galaxy populations, we find that there is no discernible difference in either the X-ray emission or other physical properties of galaxies that show \heii.

\begin{figure}
	\centering
	\includegraphics[scale=0.44]{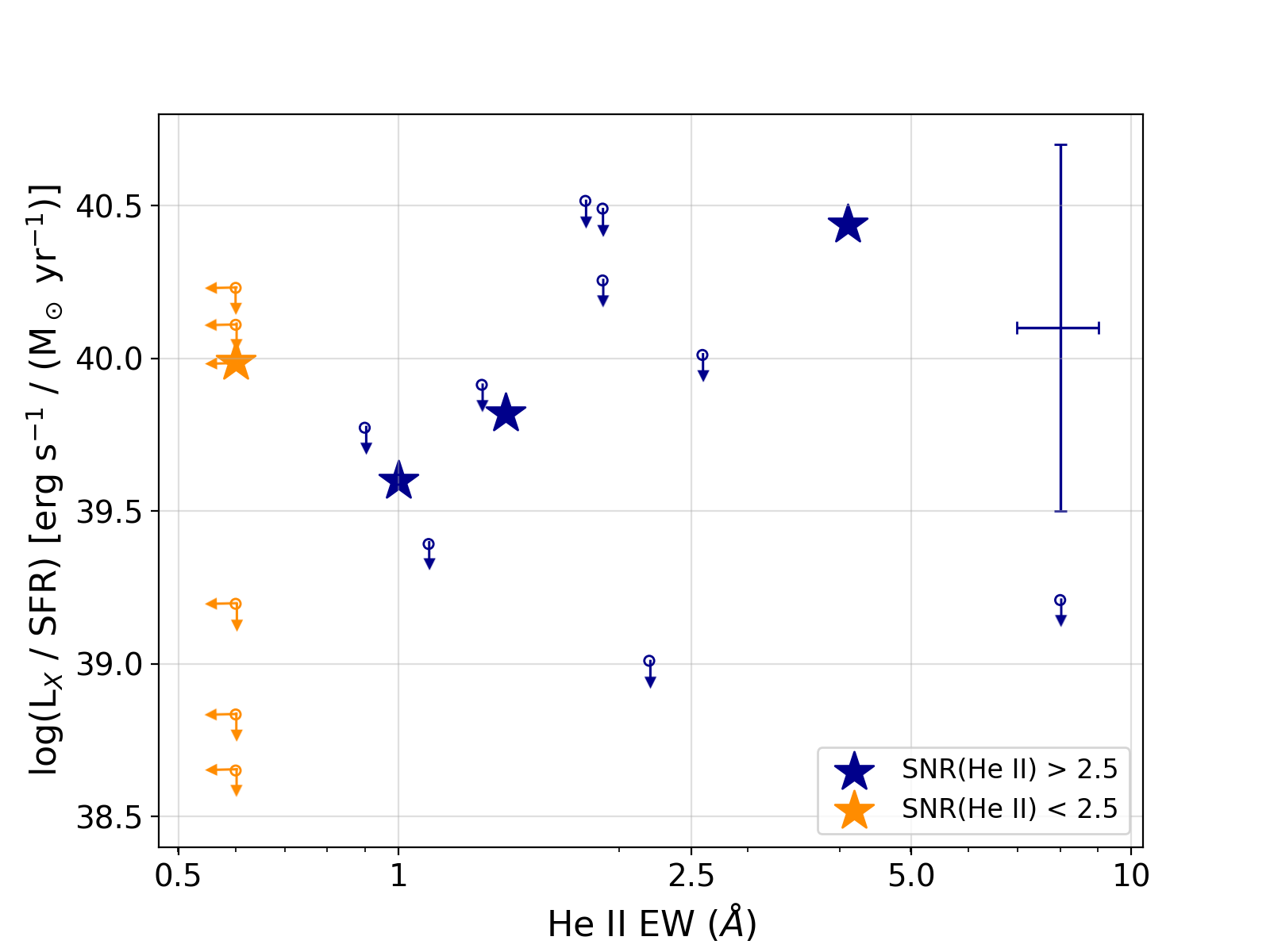}
	\caption{$L_X$/SFR as a function of \heii EW for both \emph{Bright} (blue) and \emph{Faint} (orange) \heii emitters from the sample of \citetalias{sax20}. The individual galaxies with S/N $> 2$ in the X-ray image are marked with stars and $2\sigma$ limits are marked using downward pointing arrows. The average errors on $L_X$/SFR and \heii EWs are shown in the top-right. It is clear that for \heii emission with comparable EWs, the measured $L_X$/SFR spans more than an order of magnitude. This suggests that there is no clear correlation between $L_X$/SFR and \heii EW, leading to the conclusion that XRBs are not universally dominant in \heii emitting galaxies.}
	\label{fig:xray-heii}
\end{figure}

We can also test whether there is any correlation between the strength of \heii emission line and the X-ray luminosity of individually detected sources by exploring whether $L_X$/SFR correlates with the observed EW of the \heii emission line. Shown in Figure \ref{fig:xray-heii} are $L_X$/SFR measurements and limits for all individual \heii emitters. We colour code the sources, with \emph{Bright} (S/N (\heii) $> 2.5$) \heii emitters shown in blue and \emph{Faint} (S/N (\heii) $< 2.5$) \heii emitters shown in orange. Those sources that have individual X-ray detections presented in Table \ref{tab:individual} are marked using stars. Although looking at only the brightest individual X-ray detections may suggest that $L_X$/SFR weakly correlates with \heii EW, the overwhelming majority of X-ray non-detected \heii emitters with comparable EWs suggests that there is no clear correlation between the strength of the \heii emission line and $L_X$/SFR measured in the galaxy. This is best highlighted by the highest EW \heii emitting galaxy not being detected in the X-ray image.

Our findings are comparable to what was reported by \citet{sen20}, who found no strong correlation between $L_X$/SFR and \heii/H$\beta$ ratios for a small sample of nearby galaxies either, leading them to conclude that high-mass XRBs are not the dominant sources of \heii ionising photon production. However, our results appear to be inconsistent with the findings of \citet{leb17}, \citet{sch19} and \citet{hea19}, who reported that contribution from XRBs in the well-studied metal-poor galaxy in the local Universe, I Zw 18, can account for the nebular \heii ($\lambda4686$) seen in its spectrum. These studies also showed that the X-ray luminosities observed in this galaxy are in line with the metallicity-dependence of XRBs.

The lack of excess X-ray emission from galaxies that show \heii emission in their UV spectra compared to those that do not, suggests that within the scope of this analysis, we do not find evidence of high-mass XRBs being the dominant sources of \heii ionising photon production in $z\sim3$ galaxies. Although the narrow \heii emitting galaxies do show a marginal excess in X-ray emission when compared to the sample of non emitters, with the current data (and limits on errors) available we can not conclude for certain whether this excess is statistically significant. Since the CDFS 7 Ms data is the deepest X-ray data available in any extragalactic field, the step forward might be to reduce error bars on X-ray measurements from \heii emitters through the identification of a much larger sample of \heii emitting galaxies.

\subsection{Presence of obscured AGN?}
The photon energies required to ionise \heii ($E>54$ eV) are easily produced in the accretion disks of active galactic nuclei (AGN) across redshifts. However, the presence of AGN in high-redshift galaxies leads to the excitation of other emission lines that also require extremely high energy photons, in addition to brightness at radio and X-ray wavelengths. \citetalias{sax20} used radio or X-ray detections, as well as detection of \civ in emission to identify possible AGN from the sample of \heii emitters and these AGN were removed from the analysis and the stacks presented in this study (even though the presence of both \heii and \civ can be explained using some stellar models without the need for AGN). With the addition of deeper X-ray photometry, we can explore whether weaker or obscured AGN could still be present in the sample of \heii emitters.

Thanks to the choice of observing band ($0.8-3$ keV) and the redshift of our sources, the strong Fe K$\alpha$ emission line at a rest-frame energy of 6.4 keV, often associated with reflection of X-ray emission from the accretion disk of the AGN \citep[e.g.][]{lig88}, is in principle observable for our sample. \citet{ric14} showed that the Fe K$\alpha$ line is observed both in obscured and unobscured AGN. Therefore, if the \heii emission seen in the UV spectra of certain sources is indeed originating from the central AGN in galaxies, we can expect to see some contribution of the generally bright Fe K$\alpha$ line at X-ray wavelengths probed in this study too. Comparing the X-ray counts and luminosities of samples of \heii emitting and non-emitting galaxies, we already showed that there is no statistically significant difference between the two. Therefore, in the context of emission from AGN, this means that there is no clear contribution from the Fe K$\alpha$ line to the X-ray luminosity of \heii emitters. In combination with the lack of other clear AGN signatures in the spectra of \heii galaxies presented in \citetalias{sax20}, we can conclude that the scenario where faint or obscured AGN are powering the \heii emission seen is unlikely.

Depending on the stacked X-ray luminosities determined for various classes of \heii emitters, we can calculate the likelihood of the presence of obscured AGN based on luminosity functions from the literature \citep[e.g.][]{air15, buc15}. \citet{vit18} extended such studies to fainter X-ray luminosities and calculated the fraction of obscured AGN both as a function of X-ray luminosity as well as redshift using the Chandra 7Ms image. \citet{vit18} showed that this fraction drops rapidly at luminosities below $L_{\textrm{2-10 keV}} = 10^{43}$ erg~s$^{-1}$ in the redshift range $z=3-6$ (however, their sample is incomplete below this luminosity limit too). The stacked X-ray luminosities measured for our \heii emitting sample are in the range $L_{\textrm{2-10 keV}} = 10^{41.4-41.5}$ erg~s$^{-1}$, and based on this tentative drop in the fraction of obscured AGNs at low X-ray luminosities, the likelihood of presence of obscured AGNs in the \heii emitting sample is considerably reduced. Further, \citet{cir19} showed that even obscured AGN at $z>2.5$ can have X-ray luminosities in excess of $L_X>10^{44}$ erg~s$^{-1}$, which is much larger than the luminosities we find for \heii emitting sources in this study.

\subsection{Other possible explanations for He \textsc{ii}}
Apart from XRBs and AGN, there may be localised high-mass star-formation occurring in certain regions of the galaxy that could be powering the \heii emission that is observed in the UV spectra. For high-redshift galaxies, the SED inferred physical properties (with limited resolution and sensitivity) tend to get averaged over the entire galaxy. Therefore, the similar metallicities and X-ray luminosities of galaxies that show \heii emission and those that do not suggest that the physical properties of both these classes of objects are largely similar. However, some differences in stellar populations are needed to explain the \heii emission in some galaxies. Since $L_X$/SFR of star-forming galaxies is dependent on metallicity (although with some scatter), it may be possible that there are localised regions of high-mass, low-metallicity star formation, possibly also hosting more XRBs, within \heii emitting galaxies. \citetalias{sax20} showed that the stacked rest-frame UV spectrum of all \heii emitting galaxies has stronger nebular emission lines when compared to the stack of non-\heii emitters, suggesting recent star-formation activity. However, the metallicities measured for both classes of objects were found to be comparable. A scenario where pockets of low-mass star-formation regions are present in a galaxy with an overall evolved stellar population, which ultimately power the \heii (and other nebular) emission cannot be ruled out. Since the X-ray luminosities measured from the CDFS image also encapsulate emission from the entire galaxy, it is impossible to study any spatial effect in the X-rays for the \heii emitting galaxies.

Unfortunately the age-metallicity degeneracy cannot be broken using the \citet{cul19} method, as it relies on an assumption of constant star-formation over a timescale of 100 Myr. Future observations of rest-frame optical lines for comparable samples of \heii emitters and non-emitters may offer accurate measurements of has phase metallicities and stellar ages, and shed some light on the underlying differences in their star-formation histories.

In the local Universe where galaxies are spatially resolved, it is possible to directly study the spatial overlap between \heii emission, regions of intense star-formation and/or X-ray point sources. \citet{keh18} studied X-ray emission from the metal-poor starburst galaxy SB0335-052E showing \heii $\lambda4686$ emission in the local Universe, reporting that the low X-ray luminosities of point sources detected within the galaxy effectively rule out significant contribution from XRBs to the \heii ionising budget, even though X-ray sources are spatially coincident with the \heii emitting regions. \citet{keh18} concluded that ionisation by single metal-free stars or binary stars with $Z\sim10^{-5}$ with a top-heavy initial-mass function in current stellar population models is the most likely explanation for the \heii emission observed in this particular galaxy. However, \citet{sch19} suggested that beaming effects on X-ray emission, which result in relatively low observed X-ray fluxes but do not rule out contribution from XRBs towards the \heii ionising budget, may offer an explanation.

As argued by \citetalias{sax20}, even though it remains unclear whether XRBs are the dominant producers of \heii ionising photons or not, binary-star models \citep{eld17} overall do a better job at producing more \heii ionising photons compared to single star models \citep[see also][]{ste16}. Recent modelling of production of ionising radiation in star-forming galaxies by \citet{pla19} showed that the highest \heii EWs are produced in low-metallicity stellar populations (both single and binary-star models) with high ionisation parameter values, $\log~U\ge -2$. However, to explain the highest \heii EWs observed in the literature, the stellar populations must have very young ages ($\log~\textrm{age/yr} < 7$). \citet{pla19} also showed that although contribution from XRBs could play a role, they may not be the dominant sources of \heii ionisation. Improvements in the predicted number of photons and inclusion of other physical phenomena associated with the evolution of massive (binary) stars, such as inclusion of massive stars whose outer envelope has been stripped due to binary interactions exposing a helium core \citep{got18, got19} may be needed to match the observed \heii EWs at high redshifts. 

It may also be possible that small pockets of metal-free, Pop III-like stars exist within galaxies that show that strong \heii emission \citep{tum01, sch03, sca03}. Pop III stars, in combination with a more widespread population of Pop II (metal-enriched) stars, may be able to explain the bright \heii emission seen in high redshift galaxies \citep[e.g.][]{vis17}. We note, however, that not all strong \heii emitters in the \citetalias{sax20} sample show a strong Ly$\alpha$ emission line, which is an important requirement for ionisation by Pop III-like stars. Additionally, a population of very massive stars (VMS) at low metallicities could be capable of producing the narrow \heii emission line, primarily due to strong but slower Wolf-Rayet type stellar winds \citep{gra15}.  

Finally, fast radiative shocks are known to be capable of powering high-ionisation emission lines in local, metal-poor galaxies \citep[e.g.][]{thu05, izo12} and such shocks may also play an important role in powering the narrow \heii emission seen in star-forming galaxies at high redshifts. However, isolating the impact of radiative shocks requires using the classical BPT diagnostics \citep{bal81}, and observations of rest-frame optical emission lines of \heii emitting galaxies at high redshifts using the \emph{James Webb Space Telescope} may shed some light on the effects of shocks in these galaxies.

\section{Summary and conclusions}
\label{sec:conclusions}
Building upon the sample of \heii $\lambda1640$ emitting galaxies at $z\sim2.2-5$ presented in \citetalias{sax20}, in this study we have presented their X-ray properties. We have used the \emph{Chandra} 7 Ms X-ray data in the CDFS field, which is the deepest X-ray data set available in a well studied extragalactic field. 

We have performed aperture photometry at the locations of \heii emitting galaxies to determine their X-ray fluxes. To boost the effective exposure times and infer the average X-ray properties of the population of \heii emitting galaxies, we have also employed stacking analysis to calculate stacked X-ray luminosities of the \heii emitting sample. To put the X-ray properties of \heii emitting galaxies in context, we have performed a bootstrap analysis to determine the X-ray properties of galaxies with no \heii emission in their UV spectra, but with comparable physical properties and redshifts to those that show \heii. The main conclusions of this study are as follows:
\begin{itemize}
	\item For individual galaxies with S/N $> 2$ in X-rays, we find luminosities in the range $L_\textrm{2-10 keV}= 3.1-5.6 \times 10^{41}$ erg~s$^{-1}$. By calculating the \heii ionising photons produced per X-ray luminosity, we find that X-ray binaries (XRBs) are not capable of fully powering the \heii emission line. 
	
	\item Using stacking analysis, we find the stacked X-ray luminosity of all 18 \heii emitters in the sample to be $L_\textrm{2-10 keV}=2.6\times 10^{41}$ erg~s$^{-1}$, and for the 13 narrow (FWHM(He \textsc{ii})$<1000$ km~s$^{-1}$) \heii emitters to be $L_\textrm{2-10 keV}= 3.1\times 10^{41}$ erg~s$^{-1}$. We then calculate the distribution of X-ray luminosities from randomly drawn samples of non-\heii emitting galaxies using bootstrapping, that are matched in numbers to the stacks of all and narrow \heii emitters. We find that although the stacked X-ray luminosity of \heii emitting galaxies is marginally higher than that of galaxies with no \heii, the difference is not statistically significant. Therefore, we find no evidence of enhanced X-ray emission in star-forming galaxies that show \heii emission in their spectra at $z\sim3$.
	
	\item To study what this result means for the impact of XRBs in \heii emitting galaxies, we compare the X-ray luminosity per star-formation rate ($L_X$/SFR) for galaxies with and without \heii. We find that $L_X$/SFR measured for stacks of \heii emitters are marginally higher than that measured for galaxies with no \heii emission, but these values are not significantly different and consistent within the error bars.
	
	\item The redshift evolution and metallicity dependence of $L_X$/SFR measured in our stacks is consistent with what has been reported in the literature at lower redshifts. Our measurements at $z\sim3$ are compatible with models predicting the redshift evolution of $L_X$/SFR based on the `X-ray main sequence' of star-forming galaxies, and we find little to no redshift evolution observed between $z\sim2-3$. The metallicity dependence of $L_X$/SFR we find for \heii emitters is consistent also consistent with little to no evolution at the lowest metallicity values.
	
	\item We find no clear correlation between $L_X$/SFR measured for individually X-ray detected \heii emitters, and the equivalent width of \heii emission seen in these galaxies. We show that most of the bright \heii emitters do not show any X-ray detection. Therefore, we conclude that there is no increased contribution from XRBs in galaxies that show \heii at $z\sim3$.
	
	\item In the light of these X-ray measurements, we discuss some additional mechanisms that could be powering \heii in some galaxies. Given the low values of $L_X$ inferred from stacks of both \heii emitters and non-emitters, we argue that even weak or obscured AGNs can be ruled out. Therefore, the \heii emission could either be powered by localised high-mass star-formation, very high mass single or binary stars with low metallicities, viewing angle effects from XRBs or radiative shocks.
\end{itemize}
	
To differentiate between the various underlying mechanisms that are possibly powering galaxies showing \heii emission at high redshifts, a multi-wavelength approach is essential. For example, access to rest-frame optical spectra with high S/N can help determine the physical properties of the stellar populations and enable more accurate metallicity measurements for \heii emitting galaxies. Follow-up observations with high-spatial resolution, both through imaging and spectroscopy, may help isolate regions of enhanced star-formation in these galaxies that could be powering the strong \heii emission lines observed. Improvements to modelling the origin of radiation from massive (binary) stars and including them in stellar population synthesis codes may also bring us closer to addressing the missing \heii ionising photons problem. Observations with upcoming facilities such as the \emph{James Webb Space Telescope} and the \emph{Extremely Large Telescope} may reveal answers to pressing questions surrounding the production of high energy photons from stars that ultimately escape from galaxies in the very early Universe and drive the process of reionisation.

\section*{Acknowledgements}
The authors thank the referee for useful comments and suggestions that improved the quality of this work. AS and LP would like to thank Fabrizio Fiore, Simonetta Puccetti, Andrea Ferrara and Roberto Maiolino for their valuable input. AS would like to thank Richard Ellis for useful discussions and suggestions. AC acknowledges the support from grant PRIN MIUR
2017-20173ML3WW\_001 and ASI n.2018-23-HH.0. This work has made extensive use of \textsc{jupyter} and \textsc{ipython} \citep{ipython}, \textsc{astropy} \citep{astropy}, \textsc{matplotlib} \citep{plt} and \textsc{topcat} \citep{topcat}. This work would not have been possible without the countless hours put in by members of the open-source developing community all around the world.

\section*{Data availability}
The data underlying this article are part of VANDELS, which is a European Southern Observatory (ESO) Public Spectroscopic Survey. The data can be accessed using the VANDELS database at \url{http://vandels.inaf.it/dr3.html}, or through the ESO archives. The code used to perform the analysis in this paper will be shared on reasonable request to the corresponding author.

%%%%%%%%%%%%%%%%%%%%%%%%%%%%%%%%%%%%%%%%%%%%%%%%%%



\bibliographystyle{mnras}
\bibliography{Xray-HeII} % if your bibtex file is called example.bib

%\begin{thebibliography}{99}
%\end{thebibliography}

%%%%%%%%%%%%%%%%%%%%%%%%%%%%%%%%%%%%%%%%%%%%%%%%%%

% Don't change these lines
\bsp	% typesetting comment
\label{lastpage}
\end{document}